\documentclass[twocolumn,prx,superscriptaddress,floatfix,longbibliography]{revtex4-2}

\usepackage{amsmath,amssymb,mathrsfs}
\usepackage{natbib}
\usepackage{tabularx}
\usepackage{amsfonts}
\usepackage[utf8]{inputenc}
\usepackage{subcaption}
\usepackage{amsmath}
\usepackage{comment}
\usepackage{bbold} 
\usepackage{hhline}
\usepackage{braket}
\usepackage{txfonts}
\usepackage{balance}
\usepackage{physics}
\usepackage{graphicx}
\usepackage{dcolumn}
\usepackage{bm}
\usepackage[unicode=true,bookmarks=true,bookmarksnumbered=false,bookmarksopen=false,breaklinks=false,pdfborder={0 0 1},backref=false,colorlinks=true,citecolor=purple]{hyperref}
\usepackage{notes2bib}
\bibnotesetup{
	note-name = ,
	use-sort-key = false
}
\def\be{\begin{equation}}
\def\ee{\end{equation}}
\def\bea{\begin{eqnarray}}
\def\eea{\end{eqnarray}}

\usepackage[x11names]{xcolor}

\begin{document}

\title{Solving optimization problems with local light shift encoding on Rydberg quantum annealers}

\author{Kapil Goswami} 
 \email{kgoswami@physnet.uni-hamburg.de}
\affiliation{%
 Zentrum f\"ur Optische Quantentechnologien, Universit\"at Hamburg,\\ Luruper Chaussee 149, 22761 Hamburg, Germany\\
}%

\author{Rick Mukherjee}%
\email{rick.mukherjee@physnet.uni-hamburg.de}
\affiliation{%
 Zentrum f\"ur Optische Quantentechnologien, Universit\"at Hamburg,\\ Luruper Chaussee 149, 22761 Hamburg, Germany\\
}%

\author{Herwig Ott}
\affiliation{%
Department of Physics and Research Center OPTIMAS, \\
Rheinland-Pf\"alzische Technische Universit\"at Kaiserslautern-Landau,
67663 Kaiserslautern, Germany \\ 
}%

\author{Peter Schmelcher}
\affiliation{%
 Zentrum f\"ur Optische Quantentechnologien, Universit\"at Hamburg,\\ Luruper Chaussee 149, 22761 Hamburg, Germany\\
}%
\affiliation{%
 The Hamburg Centre for Ultrafast Imaging, Universit\"at Hamburg,\\ Luruper Chaussee 149, 22761 Hamburg, Germany\\
}%

\date{\today}

\begin{abstract}
We provide a non-unit disk framework to solve combinatorial optimization problems such as \textit{Maximum Cut (Max-Cut)} and \textit{Maximum Independent Set (MIS)} on a Rydberg quantum annealer. Our setup consists of a many-body interacting Rydberg system where locally controllable light shifts are applied to individual qubits in order to map the graph problem onto the Ising spin model. Exploiting the flexibility that optical tweezers offer in terms of spatial arrangement, our numerical simulations implement the local-detuning protocol while globally driving the Rydberg annealer to the desired many-body ground state, which is also the solution to the optimization problem. Using optimal control methods, these solutions are obtained for prototype graphs with varying sizes at time scales well within the system lifetime and with approximation ratios close to one. The non-blockade approach facilitates the encoding of graph problems with specific topologies that can be realized in two-dimensional Rydberg configurations and is applicable to both unweighted as well as weighted graphs. A comparative analysis with fast simulated annealing is provided which highlights the advantages of our scheme in terms of system size, hardness of the graph, and the number of iterations required to converge to the solution.
\end{abstract}

\maketitle

\section{\label{Intro}Introduction}

Using quantum computing for applications requires a fault-tolerant scalable architecture \cite{shor1996fault,aharonov1997fault,preskill1998fault,gottesman1998theory,bova2021commercial}. The current so-called Noisy Intermediate Scale Quantum (NISQ) era is dictated by noisy qubits and gates with comparatively low fidelities \cite{preskill2018quantum,bharti2022noisy}. Despite these limitations, there is ongoing research in trying to find suitable problems for which the quantum approach is superior \cite{aaronson2010bqp,daley2022practical,bharti2022noisy,boixo2018characterizing}. In particular, there is a class of problems whose exact solutions are hard to obtain in polynomial time, and at best, in certain cases, there exist only approximate solutions to them. They fall under the category of NP-hard problems \cite{karp1975computational,papadimitriou1988optimization} and are regularly investigated beyond the realm of computational complexity theory in order to get better insight into the performance of quantum algorithms \cite{harrigan2021quantum,pagano2020quantum,graham_demonstration_2022}. In this work, we consider two such NP-hard problems, namely Max-Cut and MIS, which are combinatorial optimization problems that have many practical applications \cite{cho1998fast,hochbaum1998analysis,perelshtein2022practical,butenko2003maximum,scheideler2008log}.

When solving a combinatorial optimization problem with quantum systems, there are two pertinent challenges that arise: (i) the choice of quantum hardware and (ii) the choice of quantum algorithm to be implemented. An ideal quantum hardware should allow scalability with the number of qubits. Qubits of superconducting devices still suffer from noise and rely on a fixed architecture with localized connectivity \cite{devoret2004superconducting}, while trapped ions are still too sensitive to external fields to allow scalability for more than 50 qubits \cite{harter2014long,bruzewicz2019trapped}. Motivated by recent experiments where hundreds of neutral Rydberg atoms can be trapped and whose interactions can be manipulated \cite{browaeys2020many,Gal,gross2017quantum}, there is a tendency and perspective to use Rydberg platforms to solve optimization problems, which is also the focus of this work.

Very recent developments in solving optimization problems on Rydberg platforms include solving the weighted MIS graph problem on a Rydberg simulator \cite{ebadi2022quantum} and the Max-Cut problem using Rydberg gates \cite{graham_demonstration_2022} whose
approximation ratio suffers due to noisy gates. In both works, QAOA (Quantum Approximate Optimization Algorithm) approach is used to approximate the optimal solution \cite{farhi_quantum_2014, hadfield_quantum_2019}. In general, variational methods require finding optimal values for a  large number of variational parameters (initial state ansatz, choice of the mixer, number of layers), which by itself is found to be an NP-hard problem \cite{bittel2021training, bittel2022optimizing}. This can make the implementation of variational algorithms to solve a particular optimization problem fairly cumbersome. For this purpose, we propose an optimized quantum annealing protocol that is implemented on the Rydberg platform.

Another issue is the choice of encoding which is a scheme by which a real-world optimization problem is mapped onto a system of interacting qubits. The encoding scheme depends on the choice of both, hardware and algorithm. Max-Cut and MIS are optimization problems that can be represented in a QUBO form. In this representation, the objective function along with its constraints has a quadratic form with variables that take binary values \cite{barahona1988application,glover_tutorial_2019}. The advantage of formulating the problem in QUBO form is that it can naturally map to interacting spin models \cite{lucas_ising_2014}. However, it should be mentioned that depending on the type of optimization problem, this mapping is neither unique nor trivial to be physically realized \cite{kim_rydberg_2021,ebadi2022quantum,lanthaler2023rydberg,qiu_programmable_2020,lechner2015quantum}. For example, in \cite{ebadi2022quantum}, the MIS problem was encoded on an interacting spin system using the unit-disk encoding \cite{clark1990unit} that relies on the perfect implementation of the Rydberg blockade effect \cite{urban2009observation} between the atoms. Based on the unit-disk encoding, a problem graph with $N$ vertices will require $\sim 4N^2$ physical spins which can be a substantial overhead of resources \cite{nguyen_quantum_2022}. 

In this work, by taking advantage of localized light shifts on individual atoms, we encode and solve the Max-Cut/MIS problems using a non-unit disk mapping to spin models. The proposed scheme has the benefit of tackling weighted as well as unweighted graphs within the same framework and can in principle be generalized to other QUBO problems. By combining gradient and non-gradient optimal control methods, we implement the temporal evolution of the laser parameters that make the annealing process on the Rydberg platform efficient to solve optimization problems. Thus we obtain optimal solutions for different graphs (with sizes $5-15$, up to degree $5$) with varying \textit{hardness} within  $3-15 \mu s$ with an approximation ratio close to one. The \textit{hardness} parameter measures the complexity of the problem graph and serves as a convenient metric for benchmarking the performance of our optimal quantum annealing method. For graphs with similar size and complexity, we find that our protocol outperforms the simulated annealing algorithm \cite{szu1987fast} in terms of accuracy. Our encoding scheme should be applicable on any quantum platform that allows local qubit control along with global driving of the many-body system to its ground state.

We proceed as follows. In Section~\ref{Theory} we outline the theoretical description for the Rydberg simulator and introduce the Max-Cut/MIS problems. We propose a three-step scheme for non-unit disk encoding and solving graph problems using the Rydberg annealer architecture in Section~\ref{protocol}. In Section~\ref{Characterization} we discuss the tools to characterize the complexity of the problems and the quality of the solutions. Finally, in Section~\ref{Results}, we analyze our results, and detail the physical implementation of our scheme in Section~\ref{Expe}. Section~\ref{Conc} contains our conclusions and outlook.   

\section{Theory}\label{Theory}

In Sec.~\ref{Rydannealer}, we provide a brief introduction to the many-body Rydberg Hamiltonian which provides the setup for quantum annealing while Sec.~\ref{Prob_definition} defines the Max-Cut and MIS problems.

\subsection{Hamiltonian of Rydberg Simulator}\label{Rydannealer}

We consider ultracold atoms trapped in optical tweezers \cite{kaufman2021quantum,zeiher2016many}. They are described as two level systems consisting of a ground state ($\ket{g}$), and a Rydberg excited state ($\ket{e}$) that are optically coupled using a laser \cite{gaetan2009observation,low2012experimental}. The Hamiltonian expressed in the atomic basis reads as follows,
\be
\begin{split}
\hat{H}_{Ryd} = \frac{\Omega}{2} \sum_j \ket{e}_j \bra{g} +\ket{g}_j \bra{e} - \sum_j \Delta _j \ket{e}_j \bra{e} \\ 
+ \sum_{k<j} V_{kj}\qty(\ket{e}_k \bra{e} \otimes \ket{e}_j \bra{e})
\end{split}
\label{Ryd0} 
\ee
where $V_{kj} = C_6/|\mathbf{r}_j - \mathbf{r}_k|^6$ is the van der Waals ($C_6 > 0$) interaction between the pair of Rydberg atoms $\ket{e}_k$ and $\ket{e}_j$ where $C_6$ is the associated dispersion coefficient. Here $\mathbf{r}_j$ and $\mathbf{r}_k$ are the positions of the two atoms labeled as $j$ and $k$. The detuning $\Delta_j$ is the site-dependent laser parameter describing the difference between the frequency of the applied field and the natural frequency associated with the atomic transition. The Rabi frequency $\Omega$ is the global laser parameter that couples the two states and is proportional to the intensity of the driving field and the dipole moment associated with the atomic transition. Using the Pauli spin operator representation and setting $\Omega = 0$, the above equation reduces to the Ising Hamiltonian with a longitudinal field (up to a constant $C = \frac{1}{4} \sum_{j=1}^{N-1} \sum_{k=j+1}^{N} V_{jk}$) given as,
\be
\hat{H}_{Ising} = \sum_{j=1}^N  \qty(\frac{\Delta_j}{2} + \frac{1}{4} \sum_{k=1, j \neq k}^N V_{jk}) \hat{\sigma}^{z}_{j} 
+ \frac{1}{4} \sum_{j=1}^{N-1} \sum_{k=j+1}^N V_{jk} \hat{\sigma}^{z}_{j} \hat{\sigma}^{z}_{k},
\label{eq:IsingHam}
\ee
where the negative sign is absorbed in $\Delta_j$. The first term is the local longitudinal field which depends on the detuning as well as $ \sum_{k=1, j \neq k}^N V_{jk}$, while the second term corresponds to the interaction between spins. The above Hamiltonian is useful for the connection to optimization problems that can be formulated in the QUBO framework. 

\subsection{Max-Cut and MIS Problem Definitions}\label{Prob_definition}
\begin{figure*}[t]
	\includegraphics[width = \textwidth,trim={0cm 0cm 0cm 0cm},clip]{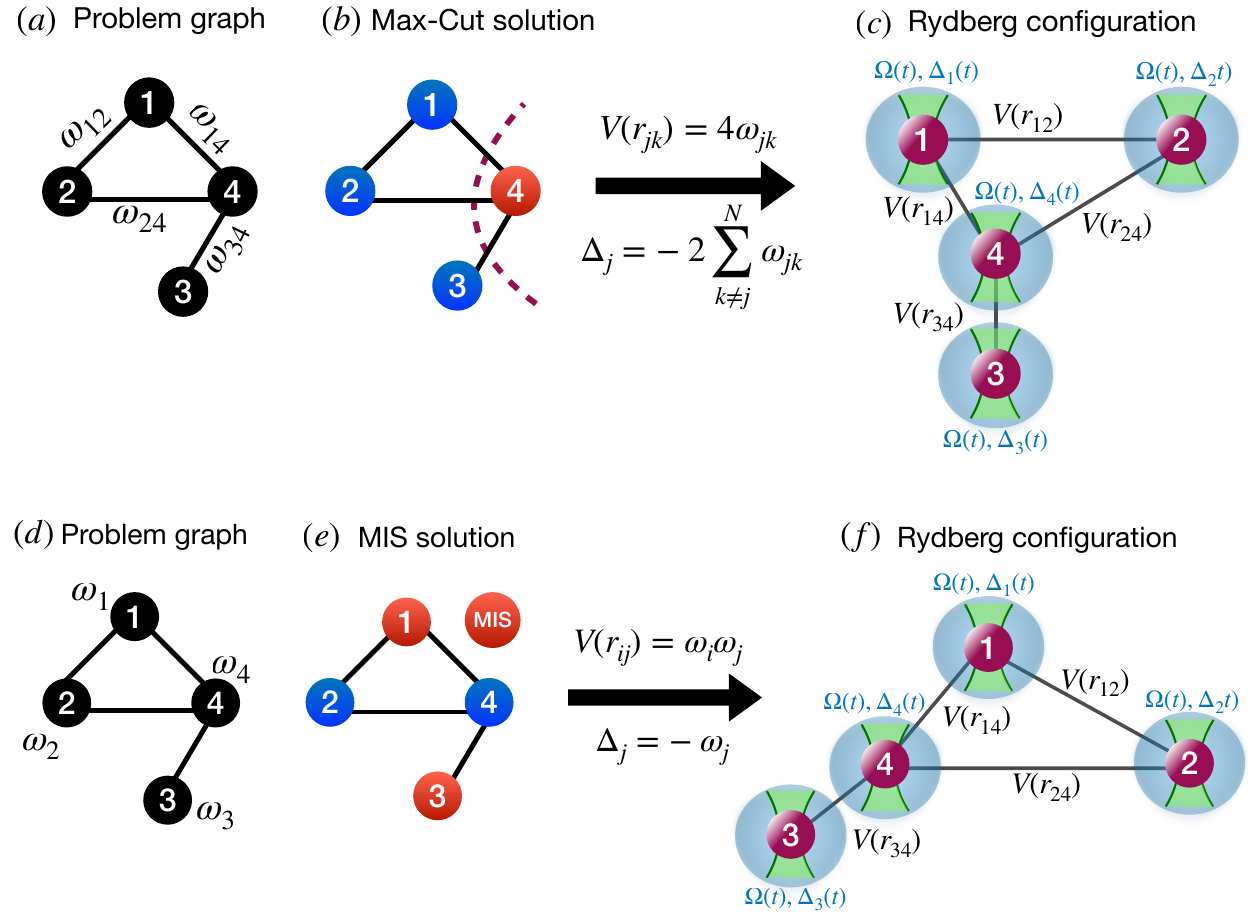}
	\caption{Setup of weighted Max-Cut and MIS problems. The figure in panel (a) is the problem graph for Max-Cut with weights $w_{ij}$ on the edges (such that $\omega_{34} > \omega_{14} > \omega_{24} >\omega_{12}$), and the figure in panel (d) is the problem graph for MIS with weights $w_i$ on the vertices (such that $\omega_{3} > \omega_{1} > \omega_{4} > \omega_{2}$). Panels (b) and (e) show the solution to the corresponding problems. The dashed curve in (b) indicates the cut, dividing the graph into two sets, red and blue vertices. In (e), red vertices constitute the MIS. Panels (c) and (f) correspond to an atomic configuration where the shaded blue region around each atom indicates the local detuning and the global Rabi frequency used in the setup. Each atom is subjected to specific values of detuning depending on the weights which are encoded in the distance-dependent interactions between the atoms as indicated with the solid-black arrow. Distance between atoms in the Rydberg configuration in the case of the Max-Cut problem follows $r_{34} < r_{14} < r_{24} < r_{12}$ and in the case of the MIS problem follows $r_{34} < r_{14} < r_{12} < r_{24}$.}
	\label{Setup}
\end{figure*}

QUBO problems are often represented by a graph $G(V, E)$ with vertices $V$ and edges $E$ which take different representations depending on the problem as shown in Fig.~\ref{Setup}. Below we define the classical cost functions for the Max-Cut and MIS problems respectively.\\

\noindent \textbf{Max-Cut:} Given a graph $G$ with weights $w_{jk}$ associated with each edge $(j,k)$, a subset of vertices from the graph is referred to as a \textit{cut}. This \textit{cut} becomes a Max-Cut when the vertices are split into two sets in a way that maximizes the total weight (Max-Cut value) of edges between them. The problem graph is illustrated in the upper panels of Fig.~\ref{Setup}(a-b). Panel (a) displays a weighted graph of size 4 and (b) shows a dashed curve representing the cut that divides the graph into two sets, blue and red. Both sets are interchangeable as the solution is degenerate. The Max-Cut cost function, which is to be maximized, is expressed as, 
\be
C_{Max-Cut} = \sum_{(j,k) \in E}w_{j,k} \qty(X_j\qty(1-X_k) + X_k\qty(1-X_j)) ,
\label{eq:MCcostM}
\ee
where the sum is over all the edges in the graph $G$. The variable $X_j \in \{0,1\}$ represents the two sets in a \textit{cut}.\\

\noindent \textbf{MIS:}  Given a graph, $G$ with weights $w_{j}$ associated with each vertex $j$, a subset of mutually non-adjacent vertices is said to be a \textit{maximum independent set} if it has the largest possible sum of weights $w_j$ over all vertices in the subset. Figure \ref{Setup} (d-e) depicts the MIS problem. MIS can be found by maximizing the following cost function,
\be
C_{MIS} = \sum_{j \in V}w_j X_j - \sum_{(j,k) \in E}w_j w_k X_j X_k ,
\label{eq:WMIScostM}
\ee
where the sum is over all the vertices $V$ in the first term and in the second term, the sum is over all the edges $E$. Here, $X_j \in \{0,1\}$ where $X_j=1$ indicates that a vertex $j$ is in the independent set. The first term in the cost function contributes only when $X_j=1$, and the second term ensures that no two vertices with an edge between them belong to the independent set. It is done by suppressing the occurrence of simultaneous $X_j=1$ and $X_k=1$ along an edge. 

\section{Non-blockade based protocol for solving Max-Cut/MIS on Rydberg annealer}\label{protocol}

This section discusses the protocol adopted in this work to solve the QUBO problems. The first step in solving the Max-Cut/MIS problems using Rydberg annealers is to encode the problem graph onto the physical spins (qubits) of the Rydberg setup. In principle, this encoding scheme is not unique and the most widely used approach, in the context of Rydberg simulators, is the unit disk (UD) encoding \cite{clark1990unit,pichler_computational_2018}. Although it is the natural choice of encoding owing to the fact that the Rydberg blockade effect successfully implements unit disks in a straight-forward manner \cite{ebadi2022quantum,nguyen_quantum_2022,kim_rydberg_2021}, it has its drawbacks. In order to solve a particular QUBO problem (in this case Max-Cut/MIS) graph with $N$ vertices, it requires solving a graph with $\sim4N^2$ unit disks. This significant overhead along with the issue of unwanted interactions calls for alternative encoding schemes. 

One of the highlights of this work is to explore an alternative, non-blockade-based encoding scheme that allows us to map QUBO to Rydberg annealers both for weighted and unweighted graphs using a single framework and it possesses a linear scaling with respect to $N$. In Steps 1 and 2, we outline our scheme of encoding the Max-Cut and MIS problem graphs onto the Rydberg spin model, while in Step 3, we discuss the implementation of optimal quantum annealing using Rydberg atoms.\\

\noindent \textbf{Step 1: Mapping of cost functions using local detuning}\\ 
The cost functions $C_{Max-Cut}$ and $C_{MIS}$  defined in the earlier section can be directly mapped to the spin Hamiltonians using the standard definitions of the Pauli operators. The details are provided in the Supplemental material \ref{steps} and the corresponding spin Hamiltonians with $N$ spins are given in the following, where
\be
\hat{H}_{Max-Cut} =\sum_{j=1}^{N-1} \sum_{k=j+1}^N w_{jk} \hat{\sigma}^{z}_{j} \hat{\sigma}^{z}_{k} 
\label{eq:Ising2M}
\ee
is the spin Hamiltonian for the Max-cut cost function Eq.(~\ref{eq:MCcostM}). Comparing Eqs. (\ref{eq:IsingHam}) and (\ref{eq:Ising2M}), we define the detunings $\Delta_j$ in a manner such that they cancel the effective longitudinal field in $\hat{H}_{Ising}$. Thus, for an atom at $j^{th}$ position, the local light shift is chosen to be $\Delta_j = - 1/2 \sum_{k \neq j}^{N} V_{jk}$. The edge weights $w_{jk}$ from the graph are identified as the interaction strength $V_{jk}$ between the atoms labeled as $j$ and $k$ as $V_{jk} = 4 w_{jk}$. This implies that the value of the detunings is related to the weights according to
\be
\Delta_j =  - 2 \sum_{k \neq j}^{N} w_{jk}.
\label{LDMC}
\ee
Similarly for the MIS problem, we get the Hamiltonian for interacting spins to be
\be 
\begin{split}
\hat{H}_{MIS} = \sum_{j=1}^N  \qty(\frac{-w_j}{2} + \frac{1}{4} \sum_{k=1, k \neq j}^N e_{jk} w_j w_k)  \hat{\sigma}_{j}^z  \\ 
+ \frac{1}{4} \sum_{j=1}^{N-1} \sum_{k=j+1}^N e_{jk} w_j w_k \hat{\sigma}_{j}^z \hat{\sigma}_{k}^z .
\end{split}
\label{eq:MISHamM}
\ee
In the case of the MIS problem, $e_{jk} = 1$, if and only if there is an edge between vertices $j$ and $k$, otherwise $e_{jk} = 0$. On comparing Eq.~(\ref{eq:IsingHam}) with (\ref{eq:MISHamM}), the vertex weights $w_{j}$ from the graph are related to the interaction strength $V_{jk}$ between the atoms labeled as $j$ and $k$ as $V_{jk} = e_{j,k} w_j w_k$. The detunings $\Delta_j$ get chosen to represent the effective longitudinal field for an atom at $j^{th}$ position. This gives the following relationship,     
\be
\Delta_j = -w_j .
\label{LDMIS2}
\ee
Thus, the mathematical problem of maximizing Eqs.~(\ref{eq:MCcostM}) and (\ref{eq:WMIScostM}) is reduced to a physical problem of finding the many-body ground state (MBGS) of the spin Hamiltonians given by Eqs.~(\ref{eq:Ising2M}) and (\ref{eq:MISHamM}) respectively. The scheme involves temporally varying the laser parameters till the specific choice of detuning values are obtained at the end of the protocol, all the while minimizing the energy of the full system. This will be elaborated in Step 3. 

The mapping of classical cost function to the Rydberg Hamiltonian described here is a \textit{direct encoding} type which results in the linear scaling of the number of atoms needed to represent the vertices of the graph. Consider a problem graph $G(\mathbf{V},\mathbf{E};\mathbf{W})$ where $n(\mathbf{V})$ and $n(\mathbf{E})$ are the number of vertices and edges respectively. Thus for an all-to-all connected graph, $n(\mathbf{E})$ takes the maximum value which is \[n(\mathbf{E_{All}}) = \frac{n(\mathbf{V})(n(\mathbf{V})-1)}{2}.\]
However, the number of edges for a general graph can vary from 1 to $n(\mathbf{E_{All}})$ depending on the problem that needs to be solved. This implies that the dependence of $n(\mathbf{E})$ on $n(\mathbf{V})$ can be expressed as 
\[n(\mathbf{E}) \propto n(\mathbf{V})^{\alpha},\] where $\alpha \in [0,2]$.  In some sense, the \textit{degrees of freedom} for a graph can be understood in terms of $n(\mathbf{E})$. Hence the number of independent degrees of freedom in $w_{jk}$ for generic graphs does not always scale quadratically with $n(\mathbf{V})$ but can have $\alpha <2$. For a direct encoding, the number of atoms required are same as the number of vertices in the problem graph. If the direct encoding of the graph with $N$ vertices were to be implemented using only interactions between atoms, then unwanted interactions cannot be ignored. This leads to unavoidably creating unwanted edges that did not exist in the original problem graph.
\\

\noindent \textbf{Step 2: Spatial arrangement of atoms}\\ 
As seen in the previous step, the distance-dependent interactions encode the information about the weights of the graph. Therefore, before any evolution of the many-body Hamiltonian, the initial arrangement of the atoms is crucial to the encoding scheme as is represented schematically in Fig \ref{Setup}. For example, in the case of a weighted Max-Cut graph with weights $\omega_{34} > \omega_{14} > \omega_{24} >\omega_{12}$ will corresponds to atoms (purple balls) arranged in a configuration such that $r_{34} < r_{14} < r_{24} < r_{12}$ which are trapped in tweezers (shown in green). Similarly, for the weighted MIS problem with weight relationship given by $\omega_{3} > \omega_{1} > \omega_{4} > \omega_{2}$ is associated with a Rydberg configuration where $r_{34} < r_{14} < r_{12} < r_{24}$. Since the weights $\omega_{ij}$ are related to the interactions $V(r_{ij})$, which in turn is reflected in the choice of the $r_{ij}$, the atoms are arranged in a manner that provides a true representation of the graph. However, it is possible to have interactions between atoms that do not share an edge in the original graph problem. Such \textit{unwanted interactions} become a serious issue for graphs with higher degrees. For this purpose, in 2D geometry, the maximum degree of a single node in the graph is limited to five for which we numerically checked that the unwanted interactions do not play a significant role. Apart from the unwanted interactions, the arrangement of the atoms which do have an edge in the problem graph is also limited. This case appears when the weights in the graph are such that the resulting relative distances between the atoms cannot be realized on a 2D plane. Although the graphs that can be represented by the geometrical arrangement of atoms are limited, we emphasize that our scheme is applicable to a large class of graphs out of which only a few are demonstrated in this work. One of the ways to generate the family of accessible graphs is by considering geometrically implementable graphs as the basis and connecting them through an additional node between them. However, adding a third dimension can also increase the value of the allowed degree and the flexibility of the atom arrangement. \\

\noindent \textbf{Step 3: Optimal Quantum Annealing with Rydberg atoms}\\ 
The previous two steps outlined the encoding of the graph problem to the Rydberg spin Hamiltonian. The goal is to find accurate solutions to a problem graph as efficiently as possible with respect to the number of iterations and run time. This is achieved by numerically solving the spin dynamics using the following Rydberg Hamiltonian for a specific arrangement of atoms
\be
\hat{H}_{Ryd} = \frac{\Omega(t)}{2} \sum_{j} \hat{\sigma}^{x}_{j} - \sum_{j} \Delta_j(t) \hat{n}^{e}_{j}
+ \sum_{k,j} V_{kj}\hat{n}^{e}_{k}\hat{n}^{e}_{j} ,
\label{Ryd1} 
\ee
where $\hat{\sigma}_{j}^x = \ket{e}_j \bra{g} +\ket{g}_j \bra{e}$  and excitation number operator $\hat{n}^{e}_{j} = \frac{1}{2} \qty(\hat{\sigma}^{z}_{j} + \mathbb{I} )$ as $\ket{e}_j \bra{e}$. The objective is to reach the target Hamiltonian $\hat{H}_{Max-Cut, MIS}$ (Eqs.~\ref{eq:Ising2M} and \ref{eq:MISHamM}) while minimizing the energy of the system to obtain the instantaneous ground state whose configuration provides the solution to the optimization problem. Initially, all atoms are in the ground state $\ket{gg...g}$ which corresponds to a large non-zero detuning value $\Delta_j(t=0) \neq 0$. During the protocol, the detuning value on each atom $\Delta_{j}(t)$ varies but attains a specific value at the end of the protocol as defined by Eqs. (\ref{LDMC}) and (\ref{LDMIS2}). The initial and final values of the Rabi frequency are set to zero. At intermediate times, non-zero values of the Rabi frequency $\Omega(t)$ provide a transverse field for the above Ising Hamiltonian. This generates quantum fluctuations and causes different many-body states to couple with each other thereby accessing a larger part of the Hilbert space. This is key to the quantum annealing process as it explores different configurations through the application of the non-zero transverse field while traversing the energy landscape to get to the desired many-body ground state configuration (i.e. the solution to the problem graph). The idea is to use well-established techniques of optimal control theory applied to quantum systems \cite{rabitz, kelly2014optimal,li2017hybrid,CRAB1, PhysRevA.91.052306, Mukherjee_2020, Mukherjee2} in order to reach this state in an efficient manner well within the system lifetime. In particular, a combination of gradient and non-gradient-based methods (see Supplemental Material \ref{OM} for details) are used to shape the pulses in time which allows us to optimally steer the many-body state towards the true solution. The initial guesses for detuning and Rabi frequency are inspired by the adiabatic evolution and are then optimized in time. The objective function that needs to be minimized during the optimal control is the expectation value of the target Hamiltonian $\hat{H}_{Max-Cut, MIS}$ with respect to the instantaneous many-body state $\ket{\psi_{inst}(t)}$ which is given as 
\be
E = \bra{\psi_{inst}(t)}\hat{H}_{Max-Cut, MIS} \ket{\psi_{inst}(t)} .
\label{expect}
\ee
$E$ is used as the cost function for the optimization but is not the typical energy of the instantaneous many-body state. Apart from $E$, we also define the overlap between the instantaneous many-body state with the many-body ground state $\ket{\psi_{g}}$ of the problem Hamiltonian $\hat{H}_{Max-Cut, MIS}$ during the protocol given as
\be
F(t) = \sum_{g} |\braket{\psi_{inst}(t)}{\psi_{g}}|^2 .
\label{fid}
\ee
The fidelity $F$ is calculated over all the degenerate ground states for an aposteriori analysis of the quality of the solution obtained. More details about the control methods are provided in Supplemental Material \ref{OM}.

\section{Characterization of Max-Cut/MIS problems} \label{Characterization}

In order to benchmark our protocol (algorithm), we need to ascertain how close is the solution provided by our algorithm to the true solution for problem graphs and compare it with other methods. This is encapsulated in a quantity referred to as \textit{approximation ratio} which is discussed in this Section. With regards to the efficiency of obtaining the solution for a given algorithm, it is possible to have a scenario where finding a solution for a particular graph is faster in run-time than compared to another graph of similar size. This indicates that the latter graph is more \textit{complex} or \textit{hard} to solve which is an inherent property of the problem graph. However, it is non-trivial to characterize the complexity of arbitrary graphs. Motivated by \cite{ebadi2022quantum,nguyen_quantum_2022}, we generalize their hardness parameter to include a broader class of problems using the notion of degenerate sub-spaces. 

\subsection{Approximation ratio} 
In general, the approximation ratio quantifies the worst-case performance of an algorithm for solving a particular problem. In certain cases, this ratio can be evaluated analytically. This is the case for the Goemans-Williamson algorithm \cite{goemans1994879} solving the Max-Cut problem which is considered to be the best classical approximate algorithm with an approximation ratio of $0.878$. Any quantum algorithm outperforming this ratio is suggestive of having an advantage over the classical case. In order to benchmark quantum algorithms, it is more convenient to evaluate this ratio numerically as defined in \cite{vazirani2001approximation} and give as
\be
R= \frac{C_{obt}}{C_{opt}} ,
\label{Approxratio}
\ee
where $C_{opt}$ is the optimal value of the cost function and $C_{obt}$ is the obtained value of the cost function evaluated through our method. In order to evaluate $C_{opt}$, the exact solution to the given problem should be available a priori. In the Results section, we evaluated $R$ for different graphs.
 \begin{table}[t]
	\begin{ruledtabular}
		\begin{tabular}{cccc} 
			\multicolumn{2}{c}{Max-Cut\footnote{For unweighted case of graph in Fig. \ref{N5} (a)}}&\multicolumn{2}{c}{MIS\footnote{For unweighted case of graph in Fig. \ref{N5} (f)}} \\ 
			\hline
			DS& MBGS& DS &MBGS \\
			\hline
			$10100$ & $\ket{egegg}$ & $10001$ & $\ket{eggge}$\\
			$01011$ & $\ket{gegee}$ & $10100$ & $\ket{egegg}$\\
			$01010$ & $\ket{gegeg}$ & $01010$ & $\ket{gegeg}$\\
			$10101$ & $\ket{egege}$ & $01001$ & $\ket{gegge}$\\
		\end{tabular}
	\end{ruledtabular}
	\caption{\label{Degen} The table shows a one-to-one correspondence between the degenerate solutions (DS) of the original Max-Cut and MIS problem to the MBGS of the corresponding Rydberg Hamiltonians given by Eq.~(\ref{eq:Ising2M}) and (\ref{eq:MISHamM}) respectively.}
\end{table}

\begin{figure*}[t]
\includegraphics[width = 1\textwidth,trim={0cm 0cm 3cm 0cm},clip]{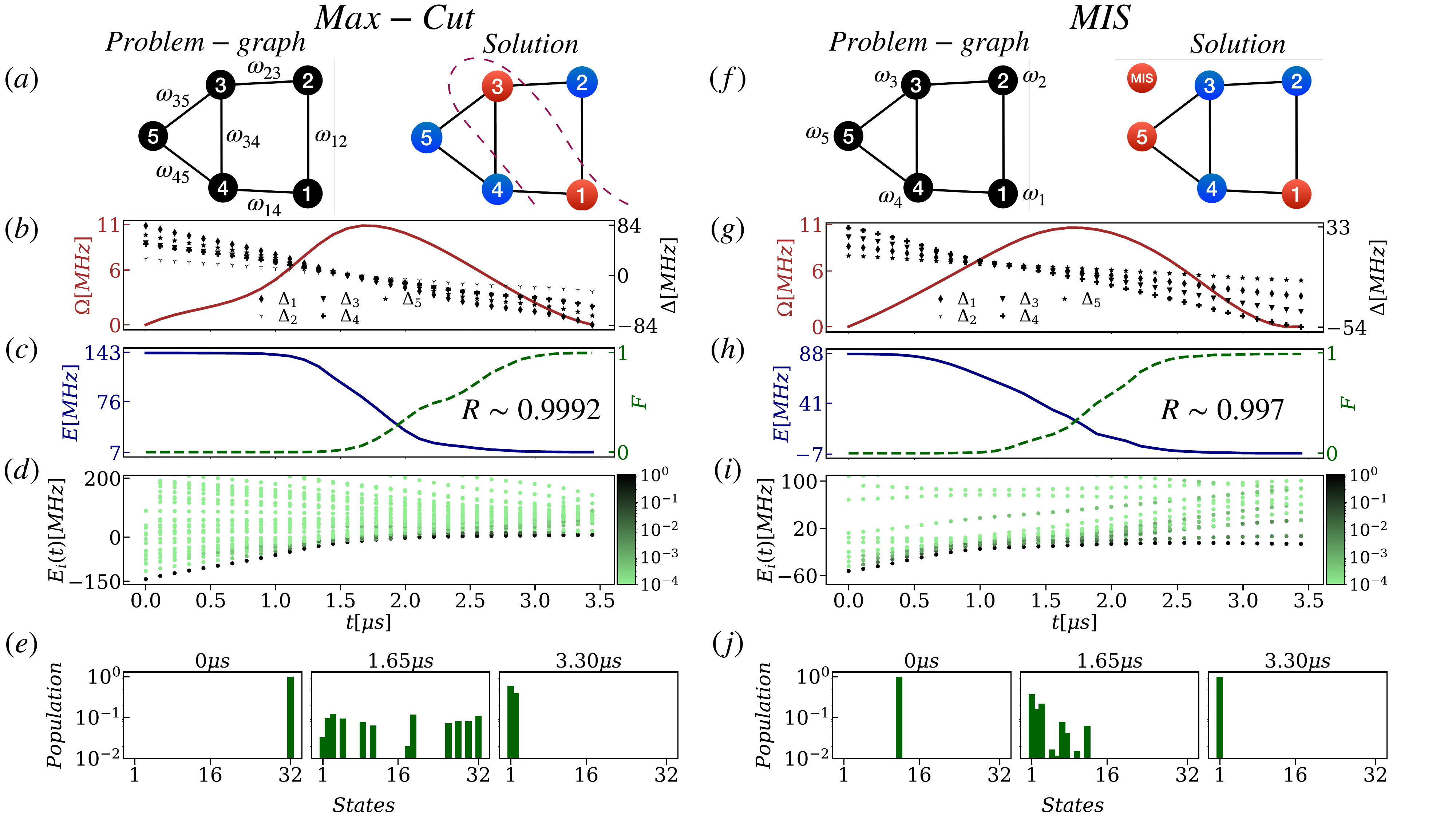}
  \caption{Left panels (a - e) are for the Max-Cut problem and right panels (f - j) are for the MIS problem. (a) and (f) show weighted prototype graphs of size $5$ with corresponding solution graphs. Panels (b) and (g) show optimal protocols for the Rabi frequency (solid dark red) and the local detunings (dotted black symbols) with time. The local detuning of each atom is controlled by varying a single time-dependent parameter $\Delta_G(t)$ which is explained in the main text. Maximum and minimum speed in detuning change for the Max-Cut problem is $47.4 MHz / \mu s$ and $15.4 MHz / \mu s $ respectively, and for the MIS problem, it is $24.5 MHz / \mu s$ and $4.7 MHz / \mu s $ respectively. The expectation value $E$ of the problem Hamiltonian with respect to the instantaneous state (solid blue) and the fidelity $F$ (dashed green) of the instantaneous state with respect to the ground state are shown in panels (c) and (h), where $R$ indicates the approximation ratio. (d) and (i) show the ordered energies of the instantaneous eigenstates, with a color bar indicating the population of the basis states during the protocol. The population of the basis states of the Hamiltonian at final time $t = T$ is shown at three different times during the protocol in (e) and (j). As shown in the middle panels of (e) and (j), $12$ states and $9$ states out of $32$ are populated in the middle of the protocol. The output at the end of the protocol captures all the degenerate states corresponding to all the degenerate Max-Cut/MIS solutions.}
  \label{N5}
\end{figure*}

\begin{figure*}[t]
\includegraphics[width = 1\textwidth,trim={0cm 0cm 0cm 0cm},clip]{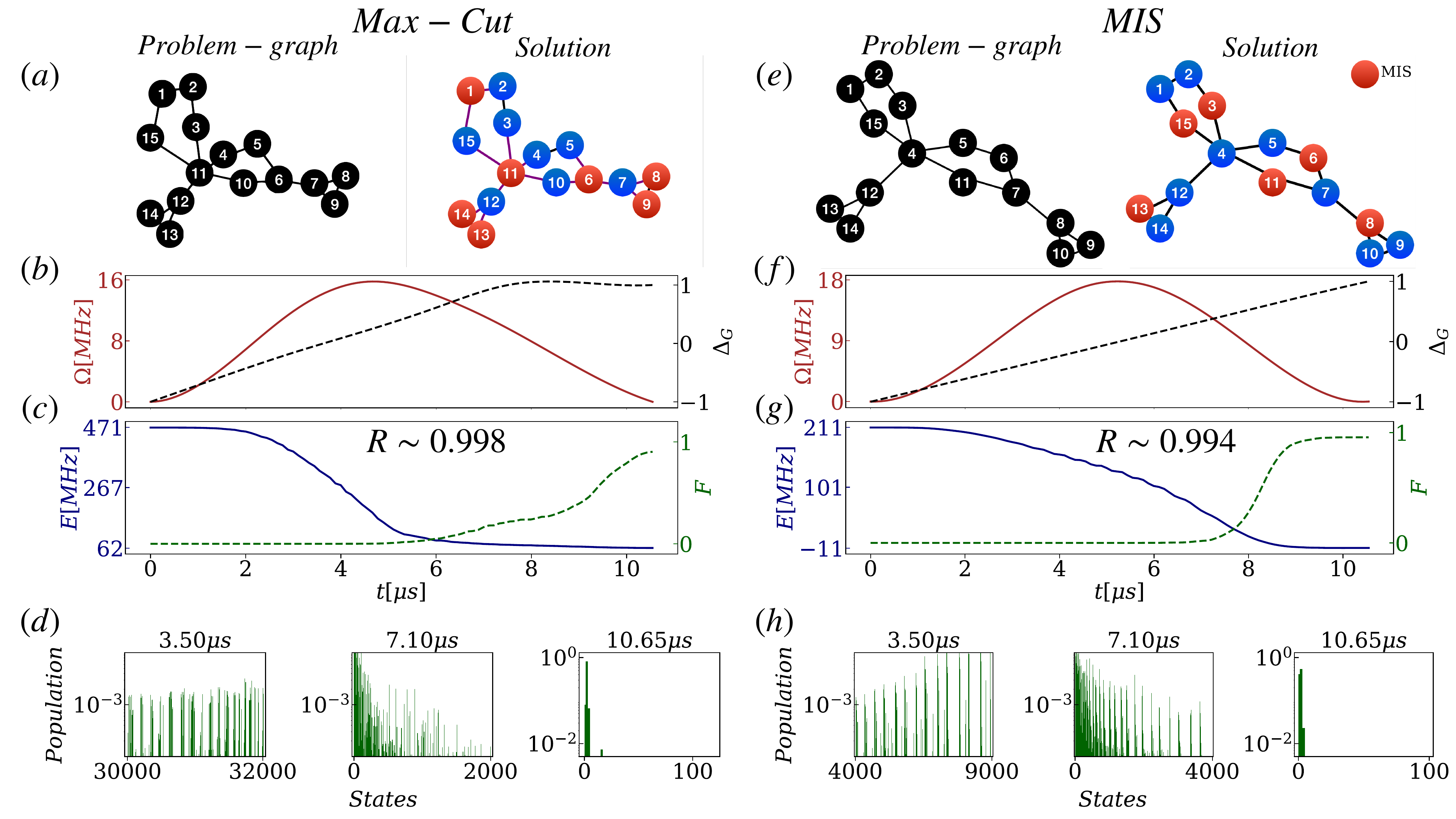}
 \caption{Solutions to the Max-Cut (a-d) and MIS (e-h) problems for a graph of size $15$ and degree $5$ using the optimal quantum annealing. Weighted prototype problem graphs with solution graphs for the Max-Cut problem are shown in (a) and are shown for the MIS problem in (e). In (a), vertex 11 has a degree of 5 while in (e), vertex 4 has a degree of 5. (b) and (f) show the optimal protocol for the Rabi frequency depicted by the solid dark red curve and $\Delta_G(t)$ (defined similar to Fig. \ref{N5}) depicted by the dashed black curve, with time. The maximum and minimum speeds for detuning change, for Max-Cut protocols, are $28.8 MHz / \mu s$ and $5.1 MHz / \mu s $ respectively, and, for MIS protocols are $1.8 MHz / \mu s$ and $0.8 MHz / \mu s $ respectively. The expectation value $E$ (solid blue) of the problem Hamiltonian with respect to the instantaneous state and fidelity $F$ (dashed green) of the instantaneous state with respect to the ground state is shown in (c) and (g), where $R$ indicates the approximation ratio. The population of states at three different times ($t=3.50 \mu s$, $t=7.10 \mu s $, and $t=10.65 \mu s$) are shown in (d,h). }
  \label{N15}
\end{figure*}

\begin{figure*}[t]
    \includegraphics[width = 1\textwidth,trim={0cm 0cm 2cm 0cm},clip]{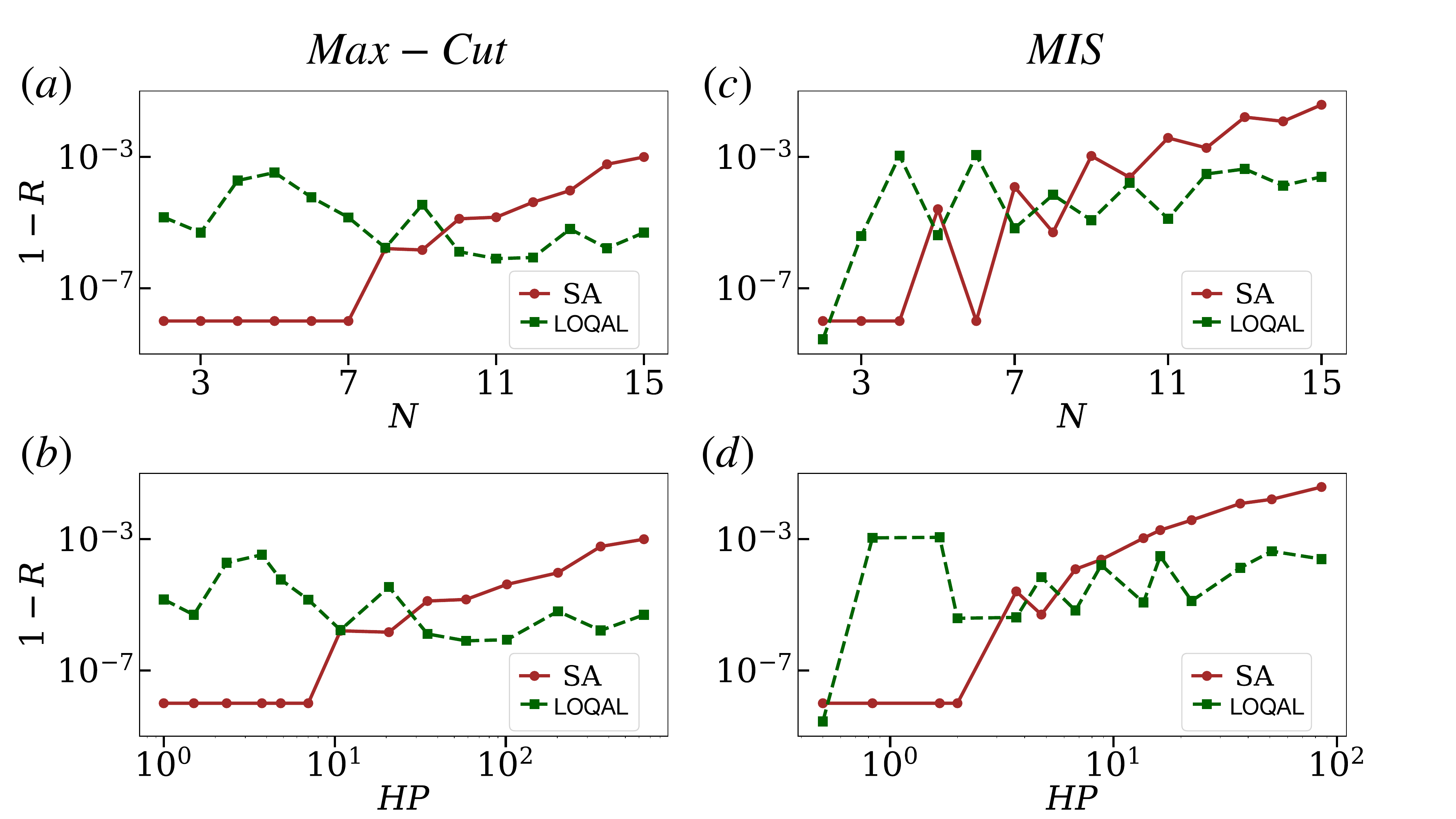}
    \caption{Comparison of optimized simulated annealing (SA) and Localised Optimal-control for Quantum Annealing in a Loop (LOQAL) for Max-Cut (a,b) and MIS (c,d) problem. Approximation ratio error $1-R$ with respect to system size $N$ is shown in (a,c), and for the hardness parameter $HP$ in (b,d).}
    \label{SAvsOQA}
\end{figure*}

\begin{figure*}[t]
\includegraphics[width = 1\textwidth,trim={4cm 0cm 2cm 0cm},clip]{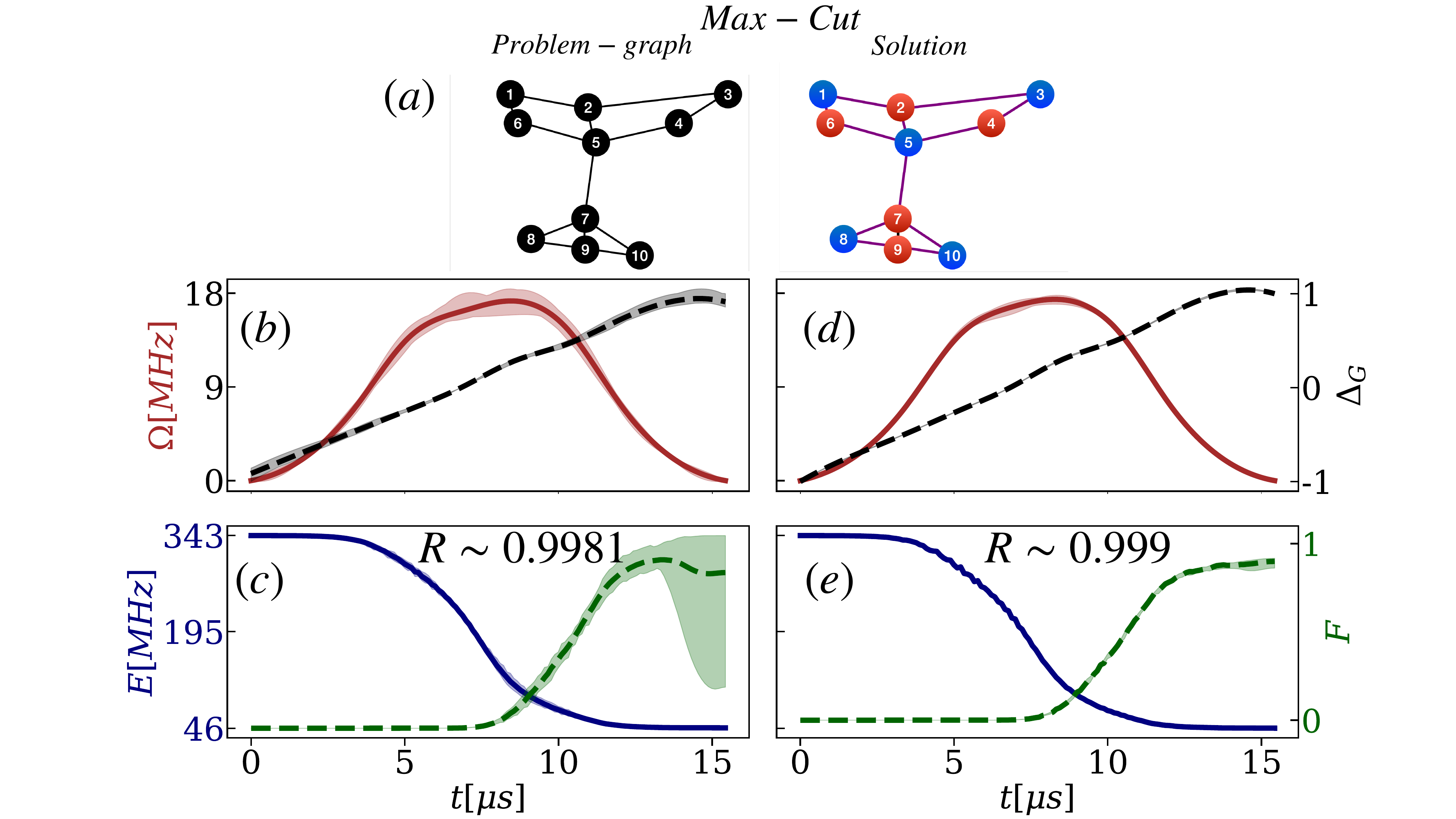}
\caption{The figure shows the results using the optimal quantum annealing with noisy protocols to solve the Max-Cut problem for a graph of size $10$. The weighted prototype graph for the Max-Cut problem is shown in (a) along with the solution graph. The optimal protocol for the Rabi frequency (solid dark red) and the factor controlling the detunings $\Delta_G$ (dashed black) are shown in (b) where noise is added to the laser parameters of the pre-optimized protocol and in (d) where noise is added during the optimization of the parameters. The shaded regions represent the fluctuations in the laser parameters for each run that were chosen from a random distribution. (c) and (e) show the corresponding expectation value (solid blue) of the problem Hamiltonian with respect to the instantaneous state ($E$) and the fidelity $F$ (dashed green) of the instantaneous state with respect to the ground state for both cases, where $R$ indicates the approximation ratio. }
  \label{N10}
\end{figure*}

\subsection{Hardness parameter} 
For Max-Cut and MIS, there are certain graphs such as weighted ones that are generally more challenging to solve computationally \cite{mohar1990eigenvalues}. One way to identify the complexity in a graph is by looking at the symmetries in the adjacency matrix \cite{ben2020symmetries}. However, a physically more intuitive way is to study the degeneracy in the many-body ground state of the interacting spin systems. As a result of mapping the cost functions using local detuning described earlier, there is a one-to-one correspondence between the degenerate solutions of the original Max-Cut and MIS problem to the many-body ground state (MBGS) of the corresponding Rydberg Hamiltonians as shown in Table~\ref{Degen}. The larger the degeneracy for the many-body ground state (solution space), the higher the probability to get to the optimal solution during the dynamics. But if the orthogonal sub-spaces (not belonging to the solution space) also have large degeneracy, then there is a possibility for the solution to get stuck in one of these unwanted sub-spaces. 

Using this notion of comparing degeneracy between relevant sub-spaces, a hardness parameter was defined only in the context of the MIS problem \cite{ebadi2022quantum}, $HP = D_{|MIS|-1}/\qty(|MIS|D_{|MIS|})$. Here $|MIS|$ and $D_{|MIS|}$ correspond to the size of the MIS and the degeneracy of the solution space. $D_{|MIS|-1}$ defines the degeneracy of the sub-space of suboptimal independent sets which has one less arbitrary element (vertex) from the MIS solution set. This hardness parameter is limited as it does not take into account all the other sub-spaces (with sizes $<|MIS|-1$) with relatively higher degeneracy. Furthermore, this hardness parameter is not well-defined for the solution to the Max-Cut problem.
 
Thus in this work, a more general hardness parameter is introduced for both Max-Cut and MIS. This parameter includes information about the degenerate sub-spaces and is defined as  
\be 
HP = \frac{\sum_{D>D_{cutoff}} D}{C_{opt} D_{opt}} , \label{HP} \ee 
where $D_{opt}$ represents the degeneracy of the solution space, $C_{opt}$ is the optimal value of the cost function and $D$ represents the degeneracy of a sub-space. The sum is taken over all orthogonal sub-spaces whose degeneracy is greater than a specific cutoff value. The cutoff value is arbitrarily chosen until the value of the hardness parameter $HP$ shows convergence. It should be noted that if $D_{opt} > D_{cutoff}$ then this degenerate configuration is excluded from the numerator sum. The hardness defined for a problem instance is an inherent property of the problem, independent of the algorithm. This is because the hardness here characterizes if one instance of the problem has a more non-convex optimization landscape as compared to the other instance. If a problem has more degenerate subspaces, the optimization landscape will be more non-convex and hence harder to navigate for any classical algorithm.
We numerically find that our hardness parameter faithfully captures the complexity of the problem graphs for various cases as shown in the Results section.

\section{\label{Results}Results}

As mentioned in Step 2 of Sec.~\ref{protocol}, although we consider prototypical graphs with a degree not more than five and with restricted connectivity that is realizable in a two-dimensional array of atoms in tweezers, we have solved for a large class of graphs out of which only a few are shown here. These other graphs can e.g. be generated by combining smaller graphs that are known to be solvable through an additional node between them. Refer to Supplemental material \ref{NM} for numerical simulation details. 

To understand the working principle of our optimal quantum annealing protocol, we consider a relatively simple case of the weighted graph of size $5$ and degree $3$ for which the Max-Cut and MIS solutions are shown in Fig. \ref{N5}(a-e) and Fig. \ref{N5}(f-j) respectively. The optimized shapes of the laser parameters (dotted black symbols for detunings, solid dark red for Rabi frequency) are shown in panels (b,g), plots of fidelity (dashed green) and energy (solid blue) in panels (c,h), plot of ordered energies $E_i(t)$ along with basis state contribution in the instantaneous eigenstates (indicated by green color bar) in panels (d,i) and the population of basis states at three different time steps in (e,j). The solutions for both problem graphs are obtained efficiently ($\sim 3.5 \mu s$) and with high fidelity ($\sim 0.99$). By turning on the Rabi pulse, one obtains multiple avoided crossings in the many-body energy spectrum. The rate at which the individual detunings are swept control the Landau-Zener transitions across multi-level crossings. This results in a non-trivial path of dynamics for the many-body system to its target state. Let $C = [\Delta_{1}(T), \Delta_{2}(T),...,\Delta_{N}(T)]$ be the set of final local detunings for $N$ atoms determined apriori by the weights on the graph given by Eqs. (\ref{LDMC}) and (\ref{LDMIS2}). The ratio between the set of detunings $\Delta_{1}(T):\Delta_{2}(T):...:\Delta_{N}(T)$ is kept fixed during the entire protocol. Thus a single time-dependent parameter $\Delta_{G} (t)$ is defined such that the temporal variation of each individual detuning  at each atom $j$ is given as $\Delta_j(t)=\Delta_{G} (t) \Delta_j (T)$ as shown in Fig. \ref{N5} (b,g). This factor $\Delta_{G} (t)$ is initialized such that it starts at a negative value at $t=0$ and increases to $1$ at $t=T$, thereby providing the desired set of detuning values at the end of the protocol. The physical implementation of the local-detuning protocol is elaborated in Sec.~\ref{Expe}. The quantity $E$ (defined by Eq.~(\ref{expect})) is minimized and this is reflected in $F$ (defined by Eq.~(\ref{fid})), shown in (c,h).  The approximation ratios $R$ (defined by Eq.~(\ref{Approxratio})) of the final state are also indicated. For a general Max-Cut problem graph, every cut has at least 2-fold degeneracy because of the symmetry between up and down spins. This is also the case for the chosen graph and thus the Max-Cut problem has more degenerate states compared to the MIS problem. This key feature is reflected in the initial part of the dynamics. The change in $E$  during the initial part of dynamics is synonymous with the change in the instantaneous eigenstates (panel (h) and (i)) for the MIS problem. However, this is not the case for the Max-Cut problem. For Max-Cut protocols, we find that $E$ stays constant for a more extended initial period signifying that the energy of the instantaneous state is close to the high energy manifold of the $\hat{H}_{Max-Cut}$ and remains there longer as a result of the degeneracy. Both for Max-Cut and MIS, the initial state $\ket{ggggg}$ is populated as shown in the left-most panel of (e) and (j). At $t=T$ for Max-Cut (right-most panel of (e)), the degenerate ground states ($\ket{egegg}$ and $\ket{gegee}$) are populated while for the MIS case, a single state $\ket{eggge}$ is populated (right-most panel of (j)). At intermediate times (middle panels of (e) and (j)) due to the non-zero values of the transverse field ($\Omega \neq 0 $), multiple states get populated. Specifically, 12 out of 32 in (e) and 9 out of 32 in (j) highlighting the fact that the optimal protocol dynamics across the energy landscape is non-intuitive. This aspect is reinforced in Fig. \ref{N15} where we have $15$ spins.

Fig. \ref{N15} demonstrates the flexibility and scaling of our encoding for a more complex case involving a weighted graph of size $15$ and  degree $5$. Panels (a) and (e) show the graph topology and the exact solution for Max-Cut and MIS. For this graph, there will be in general $15$ final detuning values $\Delta_{j}(T)$ at time $T$. The set of local detunings will all vary with the same fixed ratios between them. For illustration purposes, we show the variation of the factor $\Delta_G(t)$ in (b,f). Since the complexity of the problem is significantly increased, we have a different choice for the initial parameters. In particular, a linear guess is used for $\Delta_{G}(t)$ in (f) while a linear ramp with a flat top is used in (b). Panels (c,g) show the corresponding variation of the expectation value $E$ and fidelity $F$ during the protocol. Another signature of the complexity of this problem graph can be seen in panels (d,h) where we find the significant population of a large number of basis states out of the $2^{15}$ states, especially at inter-mediate times. In panel (d), there are $1380$ populated states at $t = 3.50 \mu s$, and $310$ populated states at $t=7.10 \mu s$. Similarly, in panel (h), there are $678$ populated states at $t =  3.50 \mu s$, and $1029$ populated states at $t = 7.10 \mu s$. It is remarkable that at the final time, we end up with a small set of degenerate MBGS. This is a direct consequence of the optimizer minimizing $E$ which forces the system to transfer the population to the lower energy levels of the target Hamiltonian. Despite the population of so many basis states, the optimal protocol ensures efficient transfer to the optimal solution ($R\sim 0.99$) making the relevance of energy gaps redundant in the context of multi-level dynamics. One may speculate that this intuition will hold for larger systems.

Fig. \ref{SAvsOQA} compares the performance of our protocol (LOQAL) with fast simulated annealing (SA) for different graphs. More details about SA are provided in the Supplemental material \ref{SAE}. In particular, we illustrate the variation of the approximation ratio error $1-R$ ($R$ is defined by Eq.~(\ref{Approxratio})) with varying system size $N = 2-15$ and hardness parameter $HP$ (defined by Eq.~(\ref{HP})). In the case of Max-Cut, the approximation ratio error ($1-R$) shows similar trends with respect to the system size $N$ and the hardness parameter $HP$ as shown in (a,b). As system size increase, so does the number of degenerate orthogonal sub-spaces (which affects the numerator of $HP$). But since the degeneracy of the solution space for the Max-Cut is always 2-fold (this affects the denominator of $HP$), the overall hardness parameter increases with system size. For the MIS problem, the SA gives a general upward trend with oscillations in the approximation ratio error with increasing system size $N$ as seen in panel (c). These oscillations are attributed to the degeneracy of the solution space being linked to the system size which by construction is 2-fold for even $N$ and 1-fold for odd $N$ for the graphs considered in this study. The larger the degeneracy of the solution space, the easier it is for the system to reach one of the solutions. This oscillatory behavior is not reflected in the hardness parameter because our definition of $HP$ takes into account the degeneracy of all the relevant subspaces and thereby successfully characterizes the problem graphs. For the graphs considered in Fig.~\ref{SAvsOQA}, it clearly shows the quantum algorithm is more robust and performs better than SA.

All results so far did not include any noise in the dynamics. Using Max-Cut as an example, we apply the optimal quantum annealing protocol to a noisy system as is expected in real experiments. Fig. \ref{N10} displays the Max-Cut problem of size $10$ with noise up to $8\%$ in the laser parameters chosen randomly which is shown as shaded regions across the bold lines (mean value) in (b,d). Two different approaches are adopted to simulate the noisy model. In (b) the noise is added after the optimization of the pulses while in (d) it is added during the optimization procedure and the optimizer adapts to the noisy protocol. The shaded region of the fidelity $F$ is broader than that of $E$ (which is barely visible) and is shown in (c), indicating that slight variations in the parameters bring a small change in $E$, but drastic changes in $F$. This is due to the fact that the gap in the lower energy levels of the target Hamiltonian is vanishing. That results in the system occupying low-energy excited states close to the MBGS, at the same time dropping the fidelity as it is a quantity highly sensitive to the occupied states. In order to better simulate real experiments, a random error of up to $8\%$ is added to the laser parameters during each run which is shown in (d). This causes the optimization landscape to adapt to the random variations in the parameters. Thus in (d), the shaded region of the pulses is not as broad as in (b). In (e) both the expectation value $E$ and the fidelity $F$ show no significant effect of the noise thereby being more resilient to the laser noise. 

\section{Physical Impementation of the protocol}\label{Expe}

For the numerical simulations shown in this work, we considered Rydberg states $60S_{1/2}$ of the Cs atoms which have a van der Waals coefficient $C_6\sim 139~GHz \cdot \mu m^6$ and a radiative lifetime of $\sim 234 \mu s$ \cite{Gal,feng2009lifetime,vsibalic2017arc}. An essential aspect of the protocol is to realize an array of trapped atoms with adjustable inter-atomic distances ranging from  $1-7\mu m$ which should be achievable with the current state-of-the-art optical tweezer technology \cite{kaufman2021quantum}. An ingredient of our protocol is the optimal control of the laser parameters. Apart from one global near-resonant laser that couples the ground state atom to its Rydberg state with Rabi frequency $\Omega$, we have another laser whose intensity will be distributed over the atoms selectively using a spatial light modulator \cite{maurer2011spatial}. This will provide individual local light shifts for the ground-Rydberg transition thereby implementing specific local detunings. At the end of the experimental sequence, measurements of the distribution of the Rydberg excitations can be performed by fluorescence imaging. The whole process will be repeated multiple times to efficiently calculate and classically minimize the expectation value of the problem Hamiltonian (Eqs.~\ref{expect}). 

We considered the case of noise in laser parameters in our simulations but the experiments can suffer from a variety of errors, introducing different types of noise to the system which includes motional dynamics, interaction induced dephasing \cite{saffman_quantum_2010},  blackbody radiation \cite{wu2022erasure} and state preparation and measurement (SPAM) errors \cite{graham2019rydberg}; all of which could lead to a significant drop in readout fidelity. There are ways to address and control errors on the Rydberg platform, including converting them into erasures \cite{wu2022erasure} and transforming complicated error models into Pauli-Z errors through the use of ancillary atoms \cite{cong2022hardware}. Future works will involve optimizing for noisy models using our protocol. \\

\section{\label{Conc}Conclusions and Outlook}

Rydberg atoms can achieve the required scalability with arbitrary connectivity between qubits making them highly desirable platforms to solve optimization problems. The protocol presented in this work is fairly universal in the sense, it can handle both weighted and unweighted graphs within the same framework which is not always obvious for other schemes and possibly could be generalized to other QUBO problems beyond Max-Cut and MIS. Although the focus of this work is on a Rydberg annealer, any quantum device that permits local qubit control along with the global driving of the many-body system can implement our protocol. A promising aspect of the optimized dynamics is that it goes beyond the energy gap dependence that would usually limit the adiabatic protocols for large system sizes, which need to be verified for larger systems.

The motivation for presenting an alternative encoding scheme that is implementable on an optimized Rydberg annealer is to have a more favorable scaling of the required number of qubits for problem graphs with $N$ vertices. For the sizes and complexity of graphs considered here, we do find accurate solutions for both, the Max-Cut and MIS problem graphs with $N=5,15$ vertices that do have a $O(N)$ scaling. However, the topology of the graphs is limited due to the issue of unwanted interactions. Such effects can be mitigated using a three-dimensional arrangement of atoms but further investigations are needed to generalize our encoding scheme for graphs with arbitrary connectivity and degree. Despite having shown the advantage of our protocol with respect to fast simulated annealing, benchmarking our method with other classical algorithms can shed more insights into the performance of our method. A more thorough analysis of optimal dynamics that can adapt to different types of noise and in particular where Bayesian methods \cite{Mukherjee_2020, Mukherjee2, Sauvage} can prove to be useful will be the scope of future work.

\begin{acknowledgments}
We are indebted to the Rymax collaboration for many fruitful and inspiring discussions, in particular with Niclas Luick, Henning Moritz, Thomas Niederpr\"um, Klaus Sengstock, and Artur Widera. This work is funded by the German Federal Ministry of Education and Research within the funding program “Quantum Technologies - from basic research to market” under Contract No. 13N16138.
\end{acknowledgments}

\bibliography{Local_detuning_Rydberg_annealer.bib}

\widetext
\clearpage

\begin{center}
\textbf{\textit{{\large Supplemental Material}}}
\end{center}
\begin{center}
\textbf{\large Solving optimization problems with local light shift encoding on Rydberg quantum annealers}
\end{center}

\setcounter{equation}{0}
\setcounter{figure}{0}
\setcounter{table}{0}
\setcounter{page}{1}
\setcounter{section}{0}
\renewcommand{\thesection}{S-\Roman{section}}
\makeatletter
\renewcommand{\theequation}{S\arabic{equation}}
\renewcommand{\thefigure}{S\arabic{figure}}

\section{Max-Cut and Maximum Independent Set encoding \label{steps}}
\noindent The mathematical mappings between the classical cost functions \cite{BOROS2002155} of Max-Cut and Maximum Independent Set to the target Hamiltonian as given in the main article are provided in this section. 

\subsection{Max-Cut}

\noindent The classical Max-Cut cost function, which is to be maximized is given by 
    \be
    C_{Max-Cut} = \sum_{(j,k) \in E}w_{jk} \qty(X_j \qty(1-X_k) + X_k(1-X_j)),
    \label{eq:MCcost}
    \ee
\noindent where $X_j \in \{ 0,1 \}$ and the sum is over all the edges $E$ with weights $W$ in the graph $G$. By replacing $X_j$ to $Z_j = 2X_j - 1$ such that $Z_j \in \{-1,1\}$, the Max-Cut cost function $C_{Max-Cut}$ becomes 
\bea 
C_{Max-Cut} &=& \sum_{(j,k) \in E}w_{jk} \qty(\frac{1+Z_j}{2}\qty(1-\frac{1+Z_k}{2}) + \frac{1+Z_k}{2}\qty(1-\frac{1+Z_j}{2})) \\
  &=& \frac{1}{4} \sum_{(j,k) \in E}w_{jk} \qty(\qty(1+Z_j) \qty(1-Z_k) + \qty(1+Z_k) \qty(1-Z_j)) \\
&=& \frac{1}{2} \sum_{(j,k) \in E}w_{jk} \qty(1 - Z_j Z_k) 
\eea

\noindent The $1/2$ in $w_{jk}$ is absorbed and the sum $\sum_{(j,k) \in E}$ over edges is changed to a double sum $\sum_{j=1}^{N-1} \sum_{k=j+1}^N$, where $N$ is the total number of vertices. This gives,

\bea 
C_{Max-Cut} &=& \sum_{j=1}^{N-1} \sum_{k=j+1}^N w_{jk} \qty(1 - Z_j Z_k),
\eea

\noindent where $w_{jk} = 0$, if there is no edge between the vertices $j$ and $k$. The above cost function can be represented as an Ising-type Hamiltonian with the help of Pauli matrices, 
    \be
     \hat{H}_I = \sum_{j=1}^{N-1} \sum_{k=j+1}^N w_{jk}  \qty(1-\hat{\sigma}^{z}_{j} \hat{\sigma}^{z}_{k}), 
    \label{eq:Isingtyp}
    \ee
\noindent such that the expectation value of $\hat{H}_I$ is the same as $C_{Max-Cut}$. The sum over the first term in $\hat{H}_I$ (\ref{eq:Isingtyp}) is just a constant, so finding the maximum $\langle \hat{H}_I\rangle$ can be formulated in terms of the Hamiltonian,  
    \be
    \hat{H}_{Max-Cut} =\sum_{j=1}^{N-1} \sum_{k=j+1}^N w_{jk} \hat{\sigma}^{z}_{j} \hat{\sigma}^{z}_{k},
    \label{eq:Ising2}
    \ee
\noindent where minimization of $\langle \hat{H}_{Max-Cut}\rangle$ leads to the maximization of $\langle \hat{H}_I\rangle$. In this way, the mathematical problem of maximizing $C_{Max-Cut}$ in Eq. (\ref{eq:MCcost}) is reduced to a physical problem of finding the many-body ground state of the Hamiltonian given by Eq. (\ref{eq:Ising2}).

\subsection{Maximum Independent Set}

\noindent The MIS can be found by maximizing the following cost function,
\be
C_{MIS} = \sum_{j \in V}w_j X_j - \sum_{(j,k) \in E}w_j w_k X_j X_k,
\label{eq:WMIScost}
\ee
\noindent where $X_j \in \{ 0,1 \}$. The sum is over all the vertices $V$ with weights $W$ in the first term and in the second term, the sum is over all the edges $E$. Similar to the case of Max-Cut, $X_j$ is replaced by $Z_j$ and the cost function $C_{MIS}$ becomes  
\bea
C_{MIS} &=& \sum_{j \in V} w_j \frac{\qty(1+Z_j)}{2} - \sum_{(j,k)\in E}w_j w_k \frac{\qty(1+Z_j)}{2} \frac{\qty(1+Z_k)}{2} \\
&=&\sum_{j \in V}\frac{w_j}{2} + \sum_{j \in V}\frac{w_j Z_j}{2} - \frac{1}{4} \sum_{(j,k)\in E} w_j w_k - \frac{1}{4} \sum_{(j,k)\in E}w_j w_k \qty(Z_j + Z_k) -\frac{1}{4} \sum_{(j,k)\in E}w_j w_k \qty(Z_j Z_k)  \\
&=& \sum_{j \in V}\frac{w_j}{2} - \frac{1}{4} \sum_{(j,k)\in E} w_j w_k + \sum_{j \in V}\frac{w_j Z_j}{2}  - \frac{1}{4} \sum_{(j,k)\in E}w_j w_k \qty(Z_j + Z_k) -\frac{1}{4} \sum_{(j,k)\in E}w_j w_k \qty(Z_j Z_k)  
\eea

\noindent Now $\sum_{(j,k)\in E} w_j w_k \qty(Z_j + Z_k)$ contains the sum over all the edges connected to the $j^{th}$ vertex. So for each vertex, the contribution is coming from the neighbors alone, 

\bea
\sum_{(j,k)\in E} w_j w_k \qty(Z_j + Z_k) = \sum_{j \in V} w_j Z_j \qty(\sum_{k \in S_j}w_k ), 
\eea
\noindent where $S_j$ is the set consisting of the neighbours of the $j^{th}$ vertex. The cost function then becomes

\bea
C_{MIS} &=& \sum_{j \in V}\frac{w_j}{2} - \frac{1}{4} \sum_{(j,k)\in E} w_j w_k + \sum_{j \in V}\frac{w_j Z_j}{2}  - \frac{1}{4} \sum_{j \in V} w_j Z_j \qty(\sum_{k \in S_j}w_k ) -\frac{1}{4} \sum_{(j,k)\in E}w_j w_k \qty(Z_j Z_k) \\
&=& \sum_{j \in V}\frac{w_j}{2} - \frac{1}{4} \sum_{(j,k)\in E} w_j w_k + \sum_{j \in V} \qty(\frac{1}{2} - \frac{\sum_{j \in S_j}w_k}{4}) w_j Z_j   -\frac{1}{4} \sum_{(j,k)\in E}w_j w_k \qty(Z_j Z_k)
\eea

\noindent The first two terms are just constants. If $Z_i$ is replaced by $\hat{\sigma}_{i}^z$, the problem of finding the maximum of C (or minimizing $-C$) is equivalent to finding the many-body ground state of $\hat{H}_{WMIS}$,
\be 
\hat{H}_{MIS} = \sum_{j \in V} \qty(\frac{\sum_{j \in S_j}w_k}{4} - \frac{1}{2} ) w_j \hat{\sigma}_{j}^z  
+ \frac{1}{4} \sum_{(j,k)\in E}w_j w_k \hat{\sigma}_{j}^z \hat{\sigma}_{k}^z,
\label{eq:MISHam0}
\ee
\noindent The sum $\sum_{(j,k) \in E} $ in the second term in Eq. (\ref{eq:MISHam0}) can be converted to a double sum as follows,

\be 
\hat{H}_{MIS} = \sum_{j=1}^N  \qty(\frac{-w_j}{2} + \frac{1}{4} \sum_{k=1, k \neq j}^N e_{jk} w_j w_k)  \hat{\sigma}_{j}^z 
+ \frac{1}{4} \sum_{j=1}^{N-1} \sum_{k=j+1}^N e_{jk} w_j w_k \hat{\sigma}_{j}^z \hat{\sigma}_{k}^z,
\label{eq:MISHam}
\ee

\noindent where $e_{ij} = 1$, if and only if there is an edge between vertices $i$ and $j$, otherwise $e_{ij} = 0$. The parameter $e_{ij}$ captures the information about the neighbors of the vertex $i$. The sum $\sum_{j \in S_i}$ is over all the neighbors of spin $i$ and is replaced by $\sum_{j=1, j \neq i}^N e_{ij}$.

\section{Numerical Details of Rydberg annealer \label{NM}}

The time-dependent Rydberg Hamiltonians provided by Eq. (\ref{Ryd1}) is implemented using the \textit{sesolve} function from QuTip library \cite{johansson2012qutip}. The objective is to find the many-body ground state of the target Hamiltonians given by Eq. (\ref{eq:Ising2M}) for Max-Cut and Eq. (\ref{eq:MISHamM}) for MIS (see main text). The expectation value $E$ of the target Hamiltonian with respect to the instantaneous many-body state is minimized during the optimal quantum annealing protocol. The profile of a global Rabi frequency $\Omega(t)$ is optimized in time while the initial and final values remain zero. An initial guess for $\Omega(t)$ is provided such that $\Omega(t) = A(1-cos(\pi t/T))^2$, where $T$ is the total time of the protocol and $A$ is of the order of few $MHz$. A localized detuning $\Delta_j(t)$ on each atom is also optimized in time where each $\Delta_j(t)$ at time $T$ is given by Eqs.~(\ref{LDMC}) and (\ref{LDMIS2}). For a particular graph, the ratios between individual $\Delta_j(T)$ are kept fixed during the entire protocol, hence, optimizing a single pre-factor $\Delta_{G}(t)$ will result in the optimization of individual detunings given by $\Delta_j(t) = \Delta_G(t)\Delta_j(T)$. $\Delta_{G}$ takes a negative value at $t=0$ such that the many-body ground state of the Hamiltonian at $t=0$ is $\ket{gg...g}$, which means all atoms are in the ground state. $\Delta_{G}$ is then varied in time and reaches one at $t=T$, the many-body ground state of the Hamiltonian at $t=T$ corresponds to the solution of Max-Cut/MIS problems. The total time $T$ is divided into sub-parts by selecting $8$ points, $\Omega(t)$ and $\Delta_{G}$ at these $8$ time points are optimized during the optimal quantum annealing to reduce $E$ and are connected by b-splines. After the system reaches its final state, fidelity $F$ is also calculated to measure the accuracy of the obtained many-body state. $F$ represents the probability of finding the system in one of the ground states, which differs from the approximation ratio. The approximation ratio $R$ is also calculated at the end of the protocol to measure the quality of the solution.

\section{Optimal Control Theory methods \label{OM}}

\noindent Optimal control theory is a mathematical framework that helps determine the best way to control a dynamical system by finding parameters that extremize a specific objective function \cite{lewis2012optimal}. It involves solving an optimization problem by adjusting control inputs over time while considering system dynamics and constraints. In physics, it has been used for shaping laser pulses, gate operations, and controlling chemical reactions \cite{werschnik2007quantum,dolde2014high,winterfeldt2008colloquium}.  For such optimization problems, gradient \cite{polak2012optimization,hasdorff1976gradient} and non-gradient \cite{conn2009introduction,hare2013survey} methods can be used. Gradient-based methods rely on calculating the gradient of the objective function with respect to the parameters and on updating them iteratively in the direction of the negative gradient to reach an optimal solution. Whereas, non-gradient methods are typically heuristic or evolutionary algorithms that iteratively explore the search space to reach an optimal solution. In this work, we use a combination of BFGS (gradient-based) \cite{broyden1970convergence,fletcher1970new,goldfarb1970family,shanno1970conditioning} and Nelder mead (non-gradient based) \cite{nelder1965simplex}. A description of both methods is given below.   
\newline

\noindent \textit{A. $-$} \textbf{Broyden-Fletcher-Goldfarb-Shanno (BFGS)} \cite{broyden1970convergence,fletcher1970new,goldfarb1970family,shanno1970conditioning} is one such gradient-based method that approximates the inverse of the Hessian matrix of the objective function using information from the gradients of the function. At each iteration, the BFGS calculates the change in the gradient and uses it to update the current estimate of the inverse Hessian. The new estimate of the inverse Hessian is then used to determine the search direction for the next iteration. The update formula is given by,  

\be
H_{k+1} = (I - \rho_k s_k y_k^T) H_k (I - \rho_k y_k s_k^T) + \rho_k s_k s_k^T
\ee

\noindent where $H_k$ is the inverse Hessian approximation at iteration $k$, $s_k = x_{k+1} - x_k$ is the difference between the current and previous estimates of the parameters, $y_k = \nabla f(x_{k+1}) - \nabla f(x_k)$ is the difference between the current and previous estimates of the gradient, $\rho_k = 1 / (y_k^T s_k)$, and $I$ is the identity matrix. The BFGS method typically starts with an initial inverse Hessian approximation, $H_0$, and iteratively updates the approximation until convergence is reached. The search direction at each iteration is given by $d_k = -H_k \nabla f(x_k)$, and the step size is determined using a line search method. 
\newline

\noindent \textit{B. $-$} As for non-gradient based, \textbf{Nelder-Mead (NM)} \cite{nelder1965simplex} is one such method, it iteratively searches for the minimum of an objective function. NM relies on exploring the simplex, a geometric figure with $n+1$ vertices in $n$ dimensions. At each iteration, the algorithm evaluates the objective function at the vertices of the simplex and then performs a set of operations to update the simplex, such as reflection, expansion, contraction, or shrinkage, based on the values of the function. The method continues until a stopping criterion is met, such as reaching a maximum number of iterations or when the function value change is small. Initialization is a step in which a simplex is defined based on an initial set of $n+1$ points in $n$-dimensional space. The function values at the vertices of the simplex are evaluated. The worst vertex of the simplex (i.e., the vertex with the highest function value) is then reflected through the centroid of the remaining vertices. If the reflected vertex has a lower function value than the second-worst vertex, the simplex is expanded in that direction. Otherwise, the reflected vertex is retained. If the reflected vertex has a higher function value than the worst vertex, the simplex is contracted towards the best vertex. If the contracted vertex has a lower function value than the worst vertex, it is retained. Otherwise, the simplex is shrunk towards the best vertex. If none of the previous steps improve the function value, the simplex is reduced toward the best vertex. Finally, the algorithm terminates when a stopping criterion is met, such as a maximum number of iterations or a slight change in the function value.
\newline

\noindent BFGS when compared to NM, can converge faster for smooth, convex functions. But at the same time, BFGS may get stuck in local minima or saddle points if the function has many local optima. Consequently, a combination of both methods is used in this work. The methods were applied sequentially in the order of BFGS-NM-BFGS (B-Nel-B), balancing exploration and exploitation. Starting with BFGS, the algorithm can quickly converge to a local minimum, and then NM is used to explore other regions of the objective function and get closer to the global optimal value. Finally, using BFGS can refine the solution and potentially converge to a better optimal value.
\newline

\noindent \textit{Parameters for NM and BFGS. $-$} Implementation in the code is done by the \textit{optimize.minimize} function from the python library SciPy \cite{virtanen2020scipy}. For both methods a convergence factor of $10^{-4}$ was set. The first layer of BFGS has a maximum iteration variable which was set to $3-6$ depending on the size of the problem and was set to $2$ for the last layer. In NM, both maximum iterations and maximum number of objective function evolution need to be fixed and were given a value of $300$ to get convergence. These values were set based on the competition between the time it takes to evaluate one layer and the quality of the solution. In the case when the system was noisy, three times more the number of runs were required as compared to the noiseless case.

\section{Simulated Annealing \label{SAE}}

\begin{figure*}[t]
\includegraphics[width = 0.8\textwidth,trim={0cm 0cm 2cm 0cm},clip]{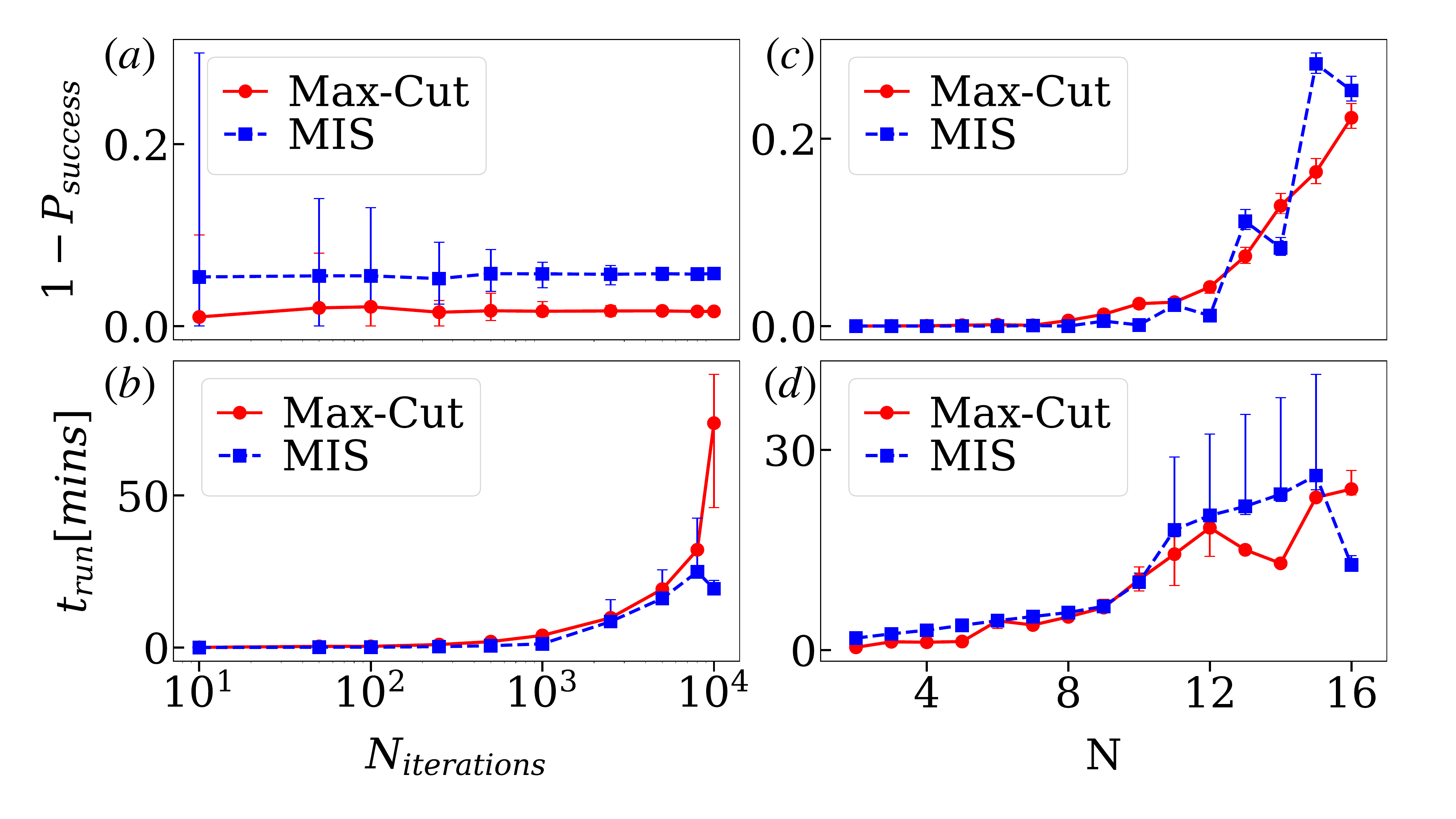}
\caption{The figure presents optimized simulated annealing for Max-Cut and MIS problems. Panel (a) shows the plot of the probability of failure ($1 - P_{success}$) and panel (b) shows run-time ($t_{run}$) in minutes as a function of the number of iterations $N_{iterations}$ (varying from $10$ to $10000$). In (a,b), a weighted problem graph with $10$ vertices and $13$ edges is chosen. (c) and (d) show the plot of the probability of failure $1 - P_{success}$ and run-time $t_{run}$ in minutes as a function of the number of vertices $N$ (varying from $2$ to $16$) for the unweighted problem graphs. }
\label{SA}
\end{figure*}

\noindent In this section, details of simulated annealing (SA) are discussed, which is used for benchmarking the quantum control method used in this work for the specific graphs. All the simulations for this method are performed by using an optimized SA from the SciPy library \cite{virtanen2020scipy}.
\newline

\noindent \textbf{Method}: Simulated annealing \cite{bertsimas1993simulated} is an algorithm used to solve optimization problems through a global search approach. It works by simulating the process of heating and cooling a material to reduce defects and minimize energy. With each function call, the algorithm searches for a new solution point in the search space. If the new point has lower energy than the previous point, it's accepted as the new optimal value with a probability of 1. If not, the probability of acceptance depends on the temperature. As the temperature decreases, the algorithm becomes more selective and only accepts better solutions. The algorithm consists of multiple cycles, each defined by a one-time cooling process to lower the temperature from an initial to a final value. While the algorithm is stochastic, it can be optimized through a local search approach to reduce the search space and find an optimal solution. Time for the procedure can further be optimized by including the annealing schedule from Fast Simulated Annealing (FSA) \cite{szu1987fast}, which consists of semi-local searches with occasional long jumps. \\

\noindent \textbf{Setting up of the numerics for SA}: For both MIS and Max-Cut problems, the cost function was defined to provide the optimum solution at the function's minima. In this work for SA, the temperature is lowered from $0.4$ to $0.01$ under a distorted Cauchy-Lorentz distribution schedule(FSA) \cite{szu1987fast,xiang1997generalized}. During the simulations, a limit of $2000$ function calls per iteration and $50$ cycles was fixed. However, the maximum number of function calls was never reached and all cycles were completed. Each full run consists of multiple iterations, an optimum is reached at the end of each full run, and $50$ such full runs were performed for statistics. The probability of success $P_{success}$ and run-time $t_{run}$ to reach the optimum is measured as a function of the number of iterations for a weighted problem graph to find an optimum number of iterations for a single run as shown in Fig \ref{SA} (a and b). $5000$ iterations were chosen for subsequent simulations, as it is enough to decrease the error in the statistics of $P_{success}$ and have a \textit{reasonable} $t_{run}$. After fixing the number of iterations, $15$ unweighted graphs were chosen of a varying number of vertices from $2$ to $16$, to study the effect of system size in finding the optimum and scaling of run-time. As shown in Fig \ref{SA} (c and d), a general trend of increase in run-time and a decrease in the probability of success is observed for both problems. An oscillatory behavior in the probability of success is also observed with varying system sizes for even and odd numbers of vertices in the graph. This behavior is attributed to the degeneracy of the solution space. If the optimum is highly degenerate, there is an increase in the probability of reaching the optimum stochastically as the solution space is larger.

\end{document}